\documentclass[twocolumn,preprintnumbers,superscriptaddress,nofootinbib]{revtex4}

\usepackage{graphicx}
\usepackage{amssymb}
\usepackage{color}
\usepackage{enumerate}

\usepackage{amssymb}
\usepackage{amsmath}
\usepackage{bm}
\usepackage{epsfig}

\begin{document}

\title{Palatini formulation of non-local gravity}
\author{F. Briscese}\email{briscese.phys@gmail.com}

\affiliation{Department of Mathematics, Physics and Electrical
Engineering, Northumbria University, Newcastle upon Tyne, UK}

\affiliation{Istituto Nazionale di Alta Matematica Francesco
Severi, Gruppo Nazionale di Fisica Matematica, Citt\`a
Universitaria, P. le A. Moro 5, 00185 Rome, EU}

\author{M. L. Pucheu}\email{mlaurapucheu@fisica.ufpb.br}
\affiliation{ Departamento de F\'isica, CCEN, Universidade Federal
da Para\'iba, Cidade Universit\'aria, 58051-970 Jo\~ao Pessoa, PB,
Brazil. }

\begin{abstract}

We derive the dynamical equations for a non-local gravity model in
the Palatini formalism and we  discuss some of the properties of
this model. We have show that, in some specific case, the vacuum
solutions of general relativity are also vacuum solutions of the
non-local model, so we conclude that, at least in this case, the
singularities of Einstein's gravity are not removed.

\end{abstract}

\maketitle

\section{introduction} \label{introduction}

Recently, the possibility of using higher-derivative theories to
construct a  viable non-local theory of quantum gravity has been
considered \cite{modesto1,biswas}. These models have been
constructed in order to fulfill the following set of hypotheses:
(1) Einstein-Hilbert action must be a good approximation of the
theory below the Planck energy scale; (2) The theory must be
perturbatively quantum-renormalizable on a flat background; (3)
The theory must be unitary; (4) Lorentz invariance must be
preserved; (5) Possibly, the presence of singularities is avoided.

Non-local models have been shown to be both renormalizable and
ghost-free \cite{modesto1,biswas} and are currently believed to be
a viable alternative to other quantum gravity scenarios, i.e. Loop
Quantum Gravity, Strings and Noncommutative Geometries (see
\cite{smolin} for a review).

The typical Lagrangian density for non-local gravity is an
extension of the Stelle theory\cite{stelle} (which is
renormalizable but plagued by ghosts) and has the following form:

\begin{equation}
\mathcal{L} =  R - \left( R_{\mu \nu}- \frac{1}{2} R g_{\mu\nu}
\right) \gamma(\Box/\Lambda^2) R^{\mu \nu} \, , \label{theory}
\end{equation}
where  the form factor $\gamma(z)$ is a non-polynomial analytic
function, $\Box$ is the covariant D'Alembertian operator and
$\Lambda$ is an invariant mass scale close to the Planck mass. The
propagator of the theory is

\begin{equation}
G(k^2) = \frac{V(k^2/\Lambda^2)  } {k^2} \left( P^{(2)} -
\frac{P^{(0)}}{2 }  \right)\,, \label{propgauge2}
\end{equation}
where $P^{(0)}$ and $P^{(2)}$ are the spin zero and spin two
projectors \footnote{In four dimensions the projectors are defined
as \cite{modesto1} $ P^{(2)}_{\mu \nu, \rho \sigma}(k) =
\frac{1}{2} ( \theta_{\mu \rho} \theta_{\nu \sigma} + \theta_{\mu
\sigma} \theta_{\nu \rho} ) - \frac{1}{3} \theta_{\mu \nu}
\theta_{\rho \sigma} \, $, $ P^{(1)}_{\mu \nu, \rho \sigma}(k) =
\frac{1}{2} \left( \theta_{\mu \rho} \omega_{\nu \sigma} +
\theta_{\mu \sigma} \omega_{\nu \rho} + \theta_{\nu \rho}
\omega_{\mu \sigma}  + \theta_{\nu \sigma} \omega_{\mu \rho}
\right) \, $, $ P^{(0)} _{\mu\nu, \rho\sigma} (k) = \frac{1}{3}
\theta_{\mu \nu} \theta_{\rho \sigma}  \, , \,\,\,\, \,\,
\bar{P}^{(0)} _{\mu\nu, \rho\sigma} (k) = \omega_{\mu \nu}
\omega_{\rho \sigma} \, $, $ \theta_{\mu \nu} = \eta_{\mu \nu} -
\frac{k_{\mu} k_{\nu}}{k^2} \, , \,\,\,\,\, \omega_{\mu \nu} =
\frac{k_{\mu} k_{\nu}}{k^2} \, $ .} and where $V(\Box/\Lambda^2)$
is defined by

\begin{equation} \gamma(\Box/\Lambda^2) \equiv
\frac{V(\Box/\Lambda^2)^{-1} -1}{\Box}  \label{FFS}  \,   .
\end{equation}

We stress that (\ref{theory}) is not the most general realization
of non-local gravity and more general examples will be considered
in the following. For instance, one may consider a model with two
form factors as in \cite{briscese1} (which has an extra scalar
degree of freedom which is not a ghost), add quadratic non-local
combinations of the Ricci tensor, etcetera. However, (\ref{theory})
is very useful to describe the key properties of non-local
gravity, so we focus on this model first.

In order to recover Einstein's gravity for small momenta, one
requires that $V(z) \simeq 1$ for $|z| \ll 1$. Moreover, if one
imposes that $V(z)$ has no poles, the theory contains only the two
massless gravitons of Einstein's theory, with no extra propagating
particles, and thus one avoids the occurrence of ghosts. Thus, if
one wants to avoid ghosts, the form factor $\gamma(z)$ cannot be
polynomial, so that the theory must contain an infinite number of
derivatives and it has to be non-local. We stress that $1/\Lambda$
represents the length scale above which the theory is fully
nonlocal and that the local behavior of the theory is recovered at
energies below $\Lambda$.

We remember that the study of non-local quantum field theory has
been introduced by Efimov in a series of seminal papers
\cite{efimovquantization,efimovunitarity,efimovcausality}, where
he discussed the quantization scheme \cite{efimovquantization},
the unitarity of the theory \cite{efimovunitarity} and the
causality \cite{efimovcausality} in the case of a non-local scalar
field (see also \cite{toboulis 2} for a recent discussion). We
also mention that a nonlocal version of QED has been studied in
\cite{efimovqed}, nonlocal vector field theory has been introduced
in \cite{tomb} and nonlocal gauge theories have been studied in
\cite{modesto nonlocal}. More recently, in \cite{biswasscalar} it
has been considered the case of a nonlocal scalar field with
specific self--interactions which have been chosen to present the
same symmetries of the non-local quantum gravity (NLQG); in
\cite{briscese de mello} the non-locality-induced one loop
corrections to the scalar field potential have been calculated and
the implications for cosmology have been discussed.

The second key property of non-local gravity, i.e. its
renormalizability, can be understood from the observation that, if
$V(z)$ goes to zero for $|z| \gg 1$ sufficiently fast, the
convergence of the propagator in the ultraviolet is improved in
such a way that the theory results to be renormalizable,
since this improves the convergence of the integrations over loops
(see for instance \cite{modestorach} for details).

At cosmological level, non-local gravity  has nice properties: in
\cite{briscese1,briscese2,modesto staro} it has been shown that on
a FRW background it reduces to the $R + \epsilon R^2$ Starobinsky
model \cite{Staro} with the identification $\epsilon \equiv
1/\Lambda^2$ up to corrections of order $1/\Lambda^4$. Therefore,
non-local gravity gives a viable inflation in agreement with
Planck data \cite{planck} for $\Lambda \sim 10^{-5} \, M_P$
\cite{briscese1,briscese2,modesto staro}.

It is notable that non-local gravity models can be constructed,
which are free from the singularities which affect Einstein's
gravity. In fact, in some specific models, the linearized
equations for gravitational perturbations of Minkowski background
typically reads $\exp(\Box/\Lambda^2) \Box h_{\mu \nu} = M_P^{-2}
\tau_{\mu \nu}$ \cite{biswas}, where $\tau_{\mu \nu}$ is the
stress energy tensor of matter. For a point-like source of mass m
this gives a Newtonian potential $h_{00} \sim erf(r\Lambda/2)\,
m/r M_P^2$, where $erf(z)$ is the error function of argument $z$,
which is finite for all $r \geq 0$. This shows how black holes
singularities of general relativity can be removed in non-local
gravity (see also \cite{modesto3}). With similar arguments one can
show that also the big bang singularity can be removed and a non
singular bouncing cosmology can be obtained \cite{modesto2}. We
stress that the disappearance of singularities is very model
dependent and for instance the specific model (\ref{theory}) still
has all the singular vacuum solutions  of Einstein's gravity
\cite{modesto singularities}, as the Schwarzschild and Kerr
metrics among others (see also \cite{modesto singularities 2} for
further discussions on non-singular spacetimes).

In this paper we are interested in deriving the Palatini
formulation of non-local gravity. The Palatini formalism
\cite{Palatini} assumes that the metric tensor and the affine
connection are independent variables, so that the field equations
are obtained by varying the action with respect to both variables.
If applied to the Einstein-Hilbert action, the Palatini method
gives the same equations of motion of general relativity but, if
one consider modifications of Einstein's gravity (see
\cite{Palatini f(R) 0} for a review of modified gravity models),
it gives a completely different theory of gravitation
\cite{Palatini f(R) 1}.

For instance, in the case of modified $f(R)$ gravity, the action
is

\begin{equation}\label{modified f(R) action}
\mathcal{S} = -\frac{1}{2 \kappa^2} \int d^4x \, \sqrt{-g} \, f(R)
+ S_m \,,
\end{equation}
where $S_m$ represents the action of all the matter fields,
$\kappa^2 = 8 \pi G$ in natural units, $G$ is the gravitational
constant and $R$ is the Ricci scalar constructed with the affine
connection $\Gamma^\alpha_{\,\, \beta \gamma}$ \footnote{Hereafter
we use the following convention: $R^{\rho}_{\,\,\, \sigma \mu
\nu}= -\partial_\nu \Gamma^\rho_{\sigma \mu} +
\partial_\mu \Gamma^\rho_{\sigma \nu} - \Gamma^\eta_{\sigma \mu}
\Gamma^\rho_{\nu \eta} + \Gamma^\eta_{\sigma \nu} \Gamma^\rho_{\mu
\eta}$, $R_{\mu \nu}=R^\alpha_{\mu \alpha \beta}$ and $R= g^{\mu
\nu} R_{\mu \nu}$. This definition differs for a minus sign from
the one used in \cite{tomboulis student}.}.  Furthermore,
henceforth we limit our analysis to a symmetric connection
$\Gamma^\alpha_{\, \, \beta \gamma}= \Gamma^\alpha_{\, \, \gamma
\beta} $.

The variation of the action (\ref{modified f(R) action}) with
respect to the metric tensor gives the set of equations

\begin{equation}\label{eq einstein palatini 1}
f'(R) R_{\mu \nu}- \frac{1}{2} f(R) g_{\mu \nu} = \kappa^2 T_{\mu
\nu}\, ,
\end{equation}
where $f'(R) \equiv \partial_R f(R)$ and $T_{\mu \nu}$ is the
energy-momentum tensor of matter, which is defined as

\begin{equation}\label{definition T}
T_{\mu \nu} \equiv  \frac{2}{\sqrt{-g}} \frac{\delta S_m}{\delta
g^{\mu \nu}} \, ,
\end{equation}
while the variation of the action (\ref{modified f(R) action})
with respect to the affine connection gives the equations
\footnote{We remember that a tensor density of weight $w$ can be
written as $t^{\mu \nu \ldots}_{\alpha \beta \ldots} \equiv
\left(-g\right)^{w/2} T^{\mu \nu \ldots}_{\alpha \beta \ldots}$,
where $T^{\mu \nu \ldots}_{\alpha \beta \ldots}$ is a tensor, and
the covariant derivative of a tensor density is defined as
$\nabla_\rho t^{\mu \nu \ldots}_{\alpha \beta \ldots} \equiv
\left(-g\right)^{w/2} \nabla_\rho T^{\mu \nu \ldots}_{\alpha \beta
\ldots}$, see \cite{MTW}. }

\begin{equation}\label{eq einstein palatini 2}
\begin{array}{ll}
 \frac{1}{2 \kappa^2}\left[\nabla_\lambda \left(\sqrt{-g} f'(R)
g^{\mu \nu} \right)- \nabla_\rho \left(\sqrt{-g} f'(R) g^{ \rho
(\mu} \right) \delta^{\nu )}_\lambda \right]= \\
\\
= -\frac{\delta S_m}{\delta \Gamma^\lambda_{\mu \nu}} \, ,
\end{array}
\end{equation}
where the second term in the l.h.s. of the last equation is
symmetrized in the indices $\mu$ and $\nu$.

In most physical cases, the matter action $S_m$ is independent
from the affine connection, thus $\delta S_m/\delta
\Gamma^\lambda_{\mu \nu} \equiv 0$ \footnote{This is not the most
general case which can be considered. For instance,  in appendix
\ref{section nonlocal scalar} it is shown that in the case of a
nonlocal scalar field one has $\delta S_m/\delta
\Gamma^\lambda_{\mu \nu} \neq 0$.}. In this case, taking the trace
on $\lambda$ and $\mu$ in Eq.(\ref{eq einstein palatini 2}), it is
evident that if $f(R) = R$, i.e. in the case of the
Einstein-Hilbert action, the affine connection is exactly the
metric connection associated to the metric tensor $g_{\mu \nu}$
and general relativity is recovered. However, when $f(R)' \neq 1$,
equation (\ref{eq einstein palatini 2}) gives the compatibility
condition

\begin{equation}\label{eq einstein palatini 2b}
\nabla_\lambda \left(f'(R) \, g^{\mu \nu} \right) = 0,
\end{equation}
and the theory is genuinely different from the corresponding
$f(R)$ model in the metric formalism. In fact,  equation (\ref{eq
einstein palatini 2b}) implies that the connection
$\Gamma^\alpha_{\beta \gamma}$ is the metric connection of the
tensor $h^{\mu\nu} \equiv f'(R) \, g^{\mu \nu}$, which is
conformal to the metric tensor $g^{\mu \nu}$.

In what follows we derive the Palatini formulation of non-local
gravity in the case of an action containing non-local quadratic
terms constructed with the Ricci scalar and with the Ricci  and
Riemann tensors. Specifically, our lagrangian density will be of
the type
\begin{equation}\label{lagrangian density}
\begin{array}{ll}
\mathcal{L}_{grav}=    \sqrt{-g}  \left[ - \frac{1}{2 \kappa^2}
\left(R + \Lambda_c\right) +
R \, h_1\left(- \square_\Lambda\right) R + \right. \\
\\
+ \left. R_{\mu \nu} \, h_2 \left(-\square_\Lambda\right) R^{\mu
\nu} + R_{\mu \nu \alpha \beta} \, h_3\left(- \square_\Lambda
\right) R^{\mu \nu \alpha \beta} \right] \, ,
\end{array}
\end{equation}
where we have included an optional cosmological constant
$\Lambda_c$ and where $\square_\Lambda \equiv \square/\Lambda^2$,
$\square \equiv g^{\mu \nu} \nabla_\mu \nabla_\nu$ is the
d'Alembertian operator constructed with the covariant derivatives
associated to the non-metric connection $\Gamma^\alpha_{\,\,\,
\beta \gamma}$ and the three form factors $h_i\left(z\right)$ are
analytic functions of their arguments and their action extend to
the objects on their right hand side, see \cite{modestorach} for
review.

Therefore the complete action for the gravitational field
$\mathcal{S}_{grav}=  \int d^4x \mathcal{L}_{grav} $ contains the
usual Einstein-Hilbert term

\begin{equation}\label{action EH}
\mathcal{S}_{EH}= - \frac{1}{2 \kappa^2} \int d^4x \sqrt{-g}
\left( R + \Lambda_c \right) \, ,
\end{equation}
plus the three contributions

\begin{equation}\label{action1}
\mathcal{S}_{Scalar}= \int d^4x \sqrt{-g} \, R \, h_1\left(-
\square_\Lambda\right) R  \, ,
\end{equation}

\begin{equation}\label{action2}
\mathcal{S}_{Ricci}= \int d^4x \sqrt{-g}  \, R_{\mu \nu} \, h_2
\left(-\square_\Lambda\right) R^{\mu \nu} \, ,
\end{equation}

\begin{equation}\label{action3}
\mathcal{S}_{Riemann}= \int d^4x \sqrt{-g}  \, R_{\mu \nu \alpha \beta}
\, h_3\left(- \square_\Lambda \right) R^{\mu \nu \alpha \beta}  \,
,
\end{equation}

The lagrangian (\ref{lagrangian density}) includes for instance,
the case of the model presented in \cite{modestorach},
corresponding to the choice $h_1 = - h_2/2$ and $h_3 =0$, which is
of particular interest since it is quantum-renormalizable and
ghost free on a flat background. We stress that the condition $h_3
=0$ does not spoil renormalizability and unitarity of the theory
around flat spacetime.

In section \ref{section action}, following the same line outlined
in \cite{modesto singularities,tomboulis student}, we derive the
field equations for the lagrangian (\ref{lagrangian density}). The
variation of the terms (\ref{action1}-\ref{action2}-\ref{action3})
are derived in sections \ref{section action 1}-\ref{section action
2}-\ref{section action 3} respectively. We stress that, since our
connection is non-metric, the operators $h_i$ do not commute with
the metric tensor $g_{\mu \nu}$ and therefore the following action
\begin{equation}\label{action4}
\mathcal{S}^*_{Ricci} =  \int d^4x \sqrt{-g}  \, R^{\mu \nu} \,
h^*_2 \left(- \square_\Lambda\right) R_{\mu \nu} \, ,
\end{equation}
is physically different from $\mathcal{S}_{Ricci}$ defined in
(\ref{action2}). Since we are interested in giving a method to
derive the equations of motion for a generic nonlocal theory
assuming the Palatini variation, in what follows we will limit our
analysis to the terms (\ref{action1}-\ref{action3}). However, with
an illustrative scope, we will calculate the variation of the
action (\ref{action4}) in Appendix \ref{section action 4}.

Then, in section (\ref{section discussion}) we will consider the
full set of equations of motion and we will show that, when
$h_3=0$, the vacuum solutions of general relativity are also
vacuum solutions of the theory (\ref{lagrangian density}), see
\cite{modesto singularities} for an analog result for the
non-local metric theory and \cite{odintsov} for the local $f(R)$
models. This fact shows that, at least in this case, the
singularities of Einstein's gravity are not removed. Finally, in
section \ref{section conclusions} we will resume the main results
of this paper and we will conclude.

\section{Equations of motion} \label{section action}

As it was mentioned, the Palatini's method does not assume
\textit{a priori} a standard form, given by the Christoffel
symbols, for the components of the affine connection. Instead, it
is based on the hypothesis that the metric tensor and the
connection are independent variables \cite{MTW} (for a review of
the history of the method see \cite{ferraris}).

If the connection is symmetric, the variation of the Riemann
tensor can be expressed as

\begin{equation}\label{delta riemann}
\delta R^\rho_{\,\,\, \sigma \mu \nu}=  \nabla_\mu \left(\delta
\Gamma^\rho_{\,\,\, \nu \sigma}\right)- \nabla_\nu \left( \delta
\Gamma^\rho_{\,\,\, \mu \sigma} \right) \,.
\end{equation}
Accordingly, for the variations of the Ricci tensor and the Ricci
scalar we have

\begin{equation}\label{delta ricci tensor}
\delta R_{\sigma  \nu}= \nabla_\rho \left(\delta
\Gamma^\rho_{\,\,\, \nu \sigma}\right)- \nabla_\nu \left( \delta
\Gamma^\rho_{\,\,\, \rho \sigma} \right)\, ,
\end{equation}

and
\begin{equation}\label{delta ricci scalar} \delta R= g^{\sigma
\nu} \delta R_{\sigma  \nu} + \delta g^{\sigma \nu}  R_{\sigma
\nu} \,.
\end{equation}

In the following, we will make extensive use of the formula of
integration by parts, which allows to show the following formula
\begin{equation}\label{action box dagger}
\begin{array}{ll}
\int d^4x \, \sqrt{-g} \, A_{\alpha_1 \alpha_2 \ldots}^{\beta_1 \beta_2\ldots} \square \, B^{\alpha_1 \alpha_2\ldots}_{\beta_1 \beta_2\ldots}= \\
\\
\int d^4x \, \sqrt{-g} \left( \square^\dagger A_{\beta_1 \beta_2 \ldots}^{\alpha_1 \alpha_2\ldots}\right) B^{\beta_1 \beta_2 \ldots}_{\alpha_1 \alpha_2\ldots} \, ,
\end{array}
\end{equation}
which is valid for any couple of tensors $A_{\alpha_1 \alpha_2\ldots}^{\beta_1 \beta_2\ldots}$ and $B^{\alpha_1 \alpha_2\ldots}_{\beta_1 \beta_2\ldots}$
with the condition that one of them is null on the boundary of integration, and where the operator $\square^\dagger $ is defined as
\begin{equation}\label{definition box dagger}
\square^\dagger T_{\alpha_1 \alpha_2...}^{\beta_1 \beta_2...}\equiv \frac{1}{\sqrt{-g}}
\nabla_\mu \nabla_\nu \left(\sqrt{-g} g^{\mu \nu} T_{\alpha_1 \alpha_2...}^{\beta_1 \beta_2...}\right) \,.
\end{equation}

Moreover, from Eq.(\ref{delta ricci tensor}), one also has

\begin{equation}\label{T delta R}
\begin{array}{ll}
\int d^4x \, \sqrt{-g} \, T^{\mu \nu} \delta R_{\mu \nu} =\\
\\
= \int d^4x  \left[\nabla_\beta \left( \sqrt{-g} \,
\delta^{(\nu}_\alpha \, T^{\mu) \, \beta} \right)  - \nabla_\alpha
\left( \sqrt{-g} T^{(\mu \nu)} \right) \right] \delta
\Gamma^\alpha_{\mu \nu} \,.
\end{array}
\end{equation}

To derive the field equations we have to calculate the functional
derivatives of the action of the gravitational field with respect
to the metric and the affine connection, which are obtained by
variation of the action with respect to such variables. The
variation of the Einstein-Hilbert term (\ref{action EH}) is
straightforward (see \cite{MTW,ferraris}), and one has

\begin{equation}\label{EoM EH 1}
\frac{\delta \mathcal{S}_{EH}}{\delta g^{\mu \nu}} = -\frac{1}{2
\kappa^2} \sqrt{-g} \left( G_{\mu \nu} - \frac{\Lambda_c}{2}
g_{\mu \nu} \right)\, ,
\end{equation}
where $G_{\mu \nu} = R_{\mu \nu}- R\,g_{\mu \nu}/2$ is the
Einstein tensor, and

\begin{equation}\label{EoM EH 2}
\frac{\delta \mathcal{S}_{EH}}{ \delta \Gamma^\alpha_{\mu \nu}} =
\frac{1}{2 \kappa^2}  \left[ \nabla_\alpha \left( \sqrt{-g} g^{\mu
\nu}\right) -  \nabla_\beta \left( \sqrt{-g} g^{\beta (\mu}
\right) \delta^{\nu)}_\alpha
 \right] \, ,
\end{equation}
where in the last term the indices $\mu$ and $\nu$ are
symmetrized. Therefore, all we need is to calculate the functional
derivatives of the terms (\ref{action1}-\ref{action3}), which is
the subject of the following sections.

\subsection{Scalar action}\label{section action 1}

To begin with, let us regard the action (\ref{action1}), which
depends on the Ricci scalar. The variation of this action is

\begin{equation}\label{delta scalar 1}
\begin{array}{ll}
\delta \mathcal{S}_{Scalar}= \int d^4x \sqrt{-g} \biggl\{ R \, \delta h_1\left(-\square_\Lambda\right)  R +\\
\\
+ R \, h_1\left(-\square_\Lambda\right)  \delta R + \delta R \, h_1\left(-\square_\Lambda\right)   R +\\
\\
-\frac{1}{2} g_{\mu \nu}  R \, h_1\left(-\square_\Lambda\right) R
\, \delta g^{\mu \nu} \biggr\} \, ,
\end{array}
\end{equation}
where we have used $\delta\sqrt{-g} = - 1/2 g_{\mu\nu} \delta
g^{\mu \nu}\sqrt{-g}$ .

To handle the first term in the r.h.s. of Eq.(\ref{delta scalar 1}) we follow the method outlined in
\cite{modesto singularities,tomboulis student}, and we expand the analytic function
$h_1\left(-\square_\Lambda\right)$ in power series as

\begin{equation} \label{definition h1}
h_1(-\square_\Lambda) = \sum^\infty_{m=0} h^{(m)}_1 \left(-\frac{\square}{\Lambda^2}\right)^m \, ,
\end{equation}
which allows to express $\delta h_1\left(-\square_\Lambda\right)$ as

\begin{equation}\label{delta h1}
\delta h_1\left(-\square_\Lambda\right)= \sum^\infty_{m=0} \sum^{m-1}_{r=0} \frac{h^{(m)}_1}{\left(-\Lambda^2\right)^m}
\square^r \delta \square \, \, \square^{m-r-1} \,.
\end{equation}
The action of $\delta \square$ on a tensor $X$ is schematically
defined by

\begin{equation} \label{delta box generic}
\left(\delta \square \right) X = \delta \left( \square X \right) - \square \delta X \,,
\end{equation}
and in the case of a scalar field $\psi$ one has
\begin{equation}\label{delta box}
\begin{array}{ll}
\left( \delta \square \right) \psi= \left[ \delta g^{\mu \nu} \nabla_\mu \nabla_\nu \psi -
g^{\mu \nu} \delta \Gamma^{\alpha}_{\mu \nu} \nabla_\alpha \psi \right]\,.
\end{array}
\end{equation}
Using Eqs.(\ref{delta h1},\ref{delta box}) one has

\begin{equation}\label{deltascalar b}
\begin{array}{ll}
\int d^4 x \sqrt{-g} R \delta h_1\left(-\square_\Lambda\right) R= - \frac{1}{\Lambda^2} \sum^\infty_{m=0} \sum^{m-1}_{r=0} h^{(m)}_1 \times \\
\\
\times \int d^4 x \sqrt{-g}
\biggl[ \left( R^{[r]^\dagger} \nabla_\mu \nabla_\nu R^{[m-r-1]} \right) \delta g^{\mu \nu}  + \\
\\
- \left(  R^{[r]^\dagger} g^{\mu \nu} \nabla_\alpha R^{[m-r-1]} \right) \delta \Gamma^\alpha_{\mu \nu} \biggr]\, ,
\end{array}
\end{equation}
where we have defined
\begin{equation} \label{eq21}
R^{[k]} \equiv \left( -\frac{\square}{\Lambda^2}\right)^k R \,,
\end{equation}
and
\begin{equation}\label{eq22}
R^{[k]^\dagger} \equiv \left( -\frac{\square^\dagger}{\Lambda^2}\right)^k R \,.
\end{equation}

Let us consider the second and third terms in Eq.(\ref{delta scalar 1}). They can be recast as

\begin{equation}\label{deltascalar d}
\begin{array}{ll}
\int d^4x \sqrt{-g} \biggl\{  R \, h_1\left(-\square/\Lambda^2\right)  \delta R + \delta R \, h_1\left(-\square/\Lambda^2\right)   R \biggr\}=  \\
\\
\int d^4x \sqrt{-g} \biggl\{ h_1\left(-\square/\Lambda^2\right) R +  h_1\left(-\square^\dagger/\Lambda^2\right)   R \biggr\} \delta R \equiv \\
\\
\equiv \int d^4x \sqrt{-g} F \delta R \, ,
\end{array}
\end{equation}
with the definition
\begin{equation}\label{dedinition F}
\begin{array}{ll}
F \equiv h_1\left(-\square/\Lambda^2\right) R +  h_1\left(-\square^\dagger/\Lambda^2\right)   R \, .
\end{array}
\end{equation}
Using Eq.(\ref{delta ricci scalar}) one also has

\begin{equation}\label{deltascalar e}
\begin{array}{ll}
\int d^4x \sqrt{-g} F \delta R = \int d^4x \sqrt{-g} F  \left[\delta g^{\mu \nu}  R_{\mu \nu} +
g^{\mu \nu} \delta R_{\mu \nu} \right]=\\
\\
= \int d^4x   \biggl\{ \sqrt{-g} \, F \,  R_{\mu \nu} \delta g^{\mu \nu} + \\
\\
+ \left[  \nabla_\alpha \left( \sqrt{-g} \,
F \, g^{\mu \nu} \right) -  \nabla_\beta \left( \sqrt{-g}\, F \, g^{\beta
(\mu} \right) \delta^{\nu)}_\alpha \right] \delta \Gamma^\alpha_{\mu \nu} \biggr\}
\, ,
\end{array}
\end{equation}
where we have used Eq.(\ref{T delta R}) with the identification $T^{\mu\nu}= F \, g^{\mu\nu} $.

Therefore, from Eqs.(\ref{delta scalar 1},\ref{deltascalar b},\ref{deltascalar e}) one has

\begin{equation}\label{functional derivative g R}
\begin{array}{ll}
\frac{\delta \mathcal{S}_{Scalar}}{\delta g^{\mu \nu}} =\sqrt{-g} \biggl\{ -\frac{1}{2} g_{\mu \nu}  R \,
h_1\left(-\square_\Lambda\right) R + F \,  R_{(\mu \nu)} +\\
\\
- \frac{1}{\Lambda^2} \sum^\infty_{m=0} \sum^{m-1}_{r=0} h^{(m)}_1
\, R^{[r]^\dagger} \nabla_{(\mu} \nabla_{\nu)} R^{[m-r-1]}
\biggr\}\, ,
\end{array}
\end{equation}
and

\begin{equation}\label{functional derivative Gamma R}
\begin{array}{ll}
\frac{\delta \mathcal{S}_{Scalar}}{\delta \Gamma^\alpha_{\mu \nu}} =
\nabla_\alpha \left( \sqrt{-g} \,
F \, g^{\mu \nu} \right) -  \nabla_\beta \left( \sqrt{-g} \, F \, g^{\beta
(\mu} \right) \delta^{\nu)}_\alpha +\\
\\
+ \frac{\sqrt{-g}}{\Lambda^2} \sum^\infty_{m=0} \sum^{m-1}_{r=0} h^{(m)}_1
\,  R^{[r]^\dagger} g^{\mu \nu} \nabla_\alpha R^{[m-r-1]} \, .
\end{array}
\end{equation}

%-------------

\subsection{Ricci action}\label{section action 2}

Let us consider in the present section, the variation of the action (\ref{action2}), which can be written as

\begin{equation}\label{variation ricci}
\begin{array}{ll}
\delta \mathcal{S}_{Ricci}= \int d^4x \sqrt{-g} \biggl\{ -\frac{1}{2} g_{\mu \nu} R_{\alpha \beta}h_2\left(-\square_\Lambda\right)R^{\alpha \beta} \delta g^{\mu \nu} +\\
\\
+  \delta R_{\mu \nu}h_2\left(-\square_\Lambda\right)R^{\mu \nu} + R_{\mu \nu} \delta h_2\left(-\square_\Lambda\right)R^{\mu \nu}+\\
\\
+ R_{\mu \nu} h_2\left(-\square_\Lambda\right) \delta R^{\mu \nu} \biggr\} \,.
\end{array}
\end{equation}

It is not difficult to see that the second and the last terms on the r.h.s of (\ref{variation ricci}) can be handled to give

\begin{equation}\label{eq30}
\begin{array}{ll}
\int d^4x \sqrt{-g} \biggl\{\delta R_{\mu \nu}h_2\left(-\square_\Lambda\right)R^{\mu \nu} +
R_{\mu \nu} h_2\left(-\square_\Lambda\right) \delta R^{\mu \nu} \biggr\}=\\
\\
\int d^4x \biggl\{ \sqrt{-g} \biggl[  R_{  \nu }^{\,\,\, \alpha}
h_2\left(-\frac{\square^\dagger}{\Lambda^2}\right)R_{ \mu \alpha}
+ R^{\alpha}_{\,\,\, \mu}
h_2\left(-\frac{\square^\dagger}{\Lambda^2}\right)R_{\alpha
\nu}\biggr] \delta g^{\mu \nu} +\\
\\
+\biggl[ \nabla_\beta \left(\sqrt{-g} \delta^{(\mu}_\alpha F^{\nu)
\beta}\right)  - \nabla_\alpha \left(\sqrt{-g}F^{(\mu \nu)}\right)
\biggr] \delta \Gamma^\alpha_{\mu \nu} \bigg\}\,,
\end{array}
\end{equation}
with $F^{\alpha \beta} \equiv h_2\left(-\frac{\square}{\Lambda^2}\right) R^{\alpha \beta} + g^{\alpha \sigma} g^{\beta \rho} h_2\left(-\frac{\square^\dagger}{\Lambda^2}\right) R_{\sigma \rho}$ and where Eq.(\ref{delta ricci tensor}) was taken into account.

Now, let us consider the following expression for the function $h_2$, analogous to the one given in (\ref{definition h1}):

\begin{equation}\label{definition h2}
h_2(-\square_\Lambda) = \sum^\infty_{m=0} h^{(m)}_2 \left(-\frac{\square}{\Lambda^2}\right)^m\,.
\end{equation}
Therefore, its consequent variation leads to
\begin{equation}\label{delta h2}
\delta h_2\left(-\square_\Lambda\right)= \sum^\infty_{m=0} \sum^{m-1}_{r=0} \frac{h^{(m)}_2}{\left(-\Lambda^2\right)^m}
\square^r \delta \square \, \, \square^{m-r-1}\,.
\end{equation}

Defining for this case

\begin{equation} \label{eq08}
R^{[k] \alpha \beta} \equiv \left( -\frac{\square}{\Lambda^2}\right)^k R^{\alpha \beta}\,,
\end{equation}

and

\begin{equation}\label{eq09}
R^{[k]^\dagger}_{ \alpha \beta} \equiv \left( -\frac{\square^\dagger}{\Lambda^2}\right)^k R_{\alpha \beta}\,,
\end{equation}

where $\square^\dagger$ is given by Eq.(\ref{definition box dagger}), the third term on the r.h.s. of (\ref{variation ricci}) can be expressed as

\begin{equation}\label{eq11}
\begin{array}{ll}
\int d^4x \sqrt{-g} R_{\mu \nu} \delta h_2\left(-\square_\Lambda\right) R^{\mu \nu}=\\
\\
-\frac{1}{\Lambda^2}\sum^{\infty}_{m=0} \sum^{m-1}_{r=0} h^{(m)}_2 \int d^4x \sqrt{-g} R^{[r]^\dagger}_{\mu \nu} \delta \left(\square \right) R^{[m-r-1] \mu \nu}\,.
\end{array}
\end{equation}

The calculation of Eq.(\ref{eq11}) requires the following identity, which can be obtained with the aid of Eq.(\ref{delta box generic}), for an arbitrary symmetric contravariant tensor of second order:
\begin{equation}\label{eq12}
\begin{array}{ll}
\left( \delta \square\right) T^{\alpha \beta}=
 \delta g^{\mu \nu} \nabla_\mu \nabla_\nu T^{\alpha \beta} + 2 \delta \Gamma^\beta_{\nu \rho} \nabla^\nu T^{\alpha \rho} + 2 \delta \Gamma^\alpha_{\nu \rho} \nabla^\nu T^{\rho \beta} +\\
 \\
 + T^{\rho \beta} \nabla^\nu \delta \Gamma^\alpha_{\nu \rho} + T^{\alpha \rho} \nabla^\nu \delta \Gamma^\beta_{\nu \rho} -g^{\mu \nu} \delta \Gamma^\lambda_{\mu \nu} \nabla_\lambda T^{\alpha \beta}\,.
\end{array}
\end{equation}

Applying this last result in (\ref{eq11}), one finally gets

\begin{equation}\label{eq13}
\begin{array}{ll}
\int d^4 x \, \sqrt{-g} \, R_{\mu \nu}\delta
h_2\left(-\square_\Lambda\right) R^{\mu \nu}=
-\frac{1}{\Lambda^2}\sum^{\infty}_{m=0} \sum^{m-1}_{r=0} h_2^{(m)}
\times\\
\\
\times \int d^4x \biggl\{ \biggl(\sqrt{-g} R^{[r]^\dagger}_{\alpha
\beta}\nabla_\mu \nabla_\nu R^{[m-r-1]\alpha \beta}\biggr)  \delta
g^{\mu\nu} + \\
\\
+ \biggl[ - \nabla_\lambda \biggl( \sqrt{-g} g^{\mu \lambda}
\left( R^{[r]^\dagger}_{\alpha \beta} R^{[m-r-1]\nu \beta}  +
R^{[r]^\dagger}_{ \beta \alpha} R^{[m-r-1]\beta \nu } \right)
\biggr) +\\
\\
+ \sqrt{-g} \biggl( 2 R^{[r]^\dagger}_{\beta \alpha}
\nabla^\mu R^{[m-r-1]\beta \nu} + 2 R^{[r]^\dagger}_{\alpha \beta} \nabla^\mu R^{[m-r-1]\nu \beta}+ \\
\\
- g^{\mu \nu} \, R^{[r]^\dagger}_{\lambda \beta} \, \nabla_\alpha
R^{[m-r-1]\lambda \beta}\biggr)  \biggr] \delta
\Gamma^{\alpha}_{\mu \nu} \biggl\}\,.
\end{array}
\end{equation}

In conclusion, the variation of the Ricci action leads to the
following equations

\begin{equation}\label{functional derivative g ricci}
\begin{array}{ll}
\frac{\delta \mathcal{S}_{Ricci}}{\delta g^{\mu \nu}}=
\sqrt{-g} \biggl\{ - \frac{1}{2}g_{\mu \nu} R_{\alpha \beta} h_2\left( -\frac{\square}{\Lambda^2} \right)R^{\alpha \beta} +\\
\\
+ R^\alpha_{ \,\,\, (\nu}
h_2\left(-\frac{\square^\dagger}{\Lambda^2}\right)R_{\alpha \mu)}+
R_{ (\nu}^{\,\,\,\,\, \alpha}
h_2\left(-\frac{\square^\dagger}{\Lambda^2}\right)R_{\mu) \alpha} +\\
\\
-\frac{1}{\Lambda^2} \sum^{\infty}_{m=0} \sum^{m-1}_{r=0}
h_2^{(m)} R^{[r]^\dagger}_{\alpha \beta} \nabla_{(\mu}
\nabla_{\nu)} R^{[m-r-1] \alpha \beta} \biggr\}\,,
\end{array}
\end{equation}

and
\begin{equation}\label{functional derivative Gamma ricci}
\begin{array}{ll}
\frac{\delta \mathcal{S}_{Ricci}}{\delta \Gamma^\alpha_{\mu \nu}}=
\nabla_\beta \left(\sqrt{-g} \delta^{(\mu}_\alpha F^{\nu) \beta}\right)  - \nabla_\alpha \left(\sqrt{-g}F^{(\mu \nu)}\right)+\\
\\
+ \frac{1}{\Lambda^2} \sum^{\infty}_{m=0} \sum^{m-1}_{r=0}
h_2^{(m)} \times \\
\\
\biggl\{   \nabla_\lambda \biggl[ \sqrt{-g} g^{ \lambda (\mu}
\left( R^{[r]^\dagger}_{\alpha \beta} R^{[m-r-1]\nu) \beta}  +
R^{[r]^\dagger}_{ \beta \alpha} R^{[m-r-1]\beta \nu) } \right)
\biggr] +\\
\\
- \sqrt{-g} \biggl[ 2 R^{[r]^\dagger}_{\beta \alpha}
\nabla^{(\mu} R^{[m-r-1]\beta \nu)} + 2 R^{[r]^\dagger}_{\alpha \beta} \nabla^{(\mu} R^{[m-r-1]\nu) \beta}+ \\
\\
- g^{\mu \nu} \, R^{[r]^\dagger}_{\lambda \beta} \, \nabla_\alpha
R^{[m-r-1]\lambda \beta}\biggr]   \biggr\}\,.
\end{array}
\end{equation}
where we have symmetrized these expressions with respect to the
indices $\mu$ and $\nu$.

\subsection{Riemann action}\label{section action 3}

In this last section, let us expand the calculation of the variation of the Riemann action (\ref{action3}). To begin with, let us consider the  expansion in power series of the function $h_3(-\square_\Lambda)$:
\begin{equation}\label{definition h3}
h_3(-\square_\Lambda) = \sum^\infty_{m=0} h^{(m)}_3 \left(-\frac{\square}{\Lambda^2}\right)^m\,.
\end{equation}

After some simple manipulation, considering (\ref{definition h3}) and (\ref{action box dagger}), the variation of (\ref{action3}) can be written as

\begin{equation}\label{variation riemann}
\begin{array}{ll}
\delta \mathcal{S}_{Riemann}= \int d^4x \sqrt{-g} \biggl\{ \biggl[  3 {{R^\pi}_\mu}^{\alpha \beta} \biggl( h_3 \bigl(-\frac{\square^\dagger}{\Lambda^2} \bigr) R_{\pi \nu \alpha \beta}\biggr) +\\
\\
- g_{\sigma \mu}{R}_{\nu \omega \alpha \beta} \biggl( h_3\left(-\frac{\square}{\Lambda^2} \right) R^{\sigma \omega \alpha \beta}\biggr) - \frac{1}{2} g_{\mu \nu} R_{\pi \sigma \alpha \beta} \times \\
\\
\times \biggl( h_3\left(-\frac{\square}{\Lambda^2} \right)R^{\pi \sigma \alpha \beta}\biggr) \biggr] \delta g^{\mu \nu} +  \biggl[ g_{\mu \rho} \biggl( h_3\left(-\frac{\square}{\Lambda^2} \right)R^{\mu \nu \alpha \beta} \biggr)+\\
\\
+ g^{\mu \nu} g^{\sigma \alpha} g^{\beta \pi} \biggl( h_3\left(-\frac{\square^\dagger}{\Lambda^2} \right) R_{\rho \mu \sigma \pi}\biggr) \biggr] \delta {R^\rho}_{\nu \alpha \beta} + \\
\\
+ R_{\mu \nu \alpha \beta} \delta h_3\left(-\frac{\square}{\Lambda^2} \right) R^{\mu \nu \alpha \beta} \biggr\}\,.
\end{array}
\end{equation}

The terms in (\ref{variation riemann}) involving the variation of the Riemann tensor give

\begin{equation}\label{eq26}
\begin{array}{ll}
\int d^4x \sqrt{-g} \delta {R^\rho}_{\nu \alpha \beta} \biggl[ g_{\mu \rho}  h_3\left(-\frac{\square}{\Lambda^2} \right)R^{\mu \nu \alpha \beta}  + g^{\mu \nu} g^{\sigma \alpha} g^{\beta \pi} \times \\
\\
\times  h_3 \bigl(-\frac{\square^\dagger}{\Lambda^2} \bigr) R_{\rho \mu \sigma \pi} \biggr] = \int d^4x \biggl[ \nabla_\beta \biggl( \sqrt{-g} g^{\rho \nu} g^{\sigma \mu} g^{\beta \pi} \times \\
\\
\times h_3\left(-\frac{\square^\dagger}{\Lambda^2} \right)  R_{\alpha \rho \sigma \pi} \biggr) -  \nabla_\rho \biggr(\sqrt{-g} g_{\alpha \beta} h_3\left(-\frac{\square}{\Lambda^2} \right)  R^{\beta \nu \rho \mu} \biggr)\biggr] \delta \Gamma^\alpha_{\mu \nu}\,,
\end{array}
\end{equation}
where we have used Eq.(\ref{delta riemann}).

Analogously to the previous sections, the calculation of the last term in (\ref{variation riemann}) makes use of the expression of the variation of $h_3(-\square_\Lambda)$

\begin{equation}\label{delta h3}
\delta h_3\left(-\square_\Lambda\right)= \sum^\infty_{m=0} \sum^{m-1}_{r=0} \frac{h^{(m)}_3}{\left(-\Lambda^2\right)^m}
\square^r \delta \square \, \, \square^{m-r-1}\,,
\end{equation}
and of the following property for an arbitrary fourth rank tensor
\begin{equation}\label{eq27}
\begin{array}{ll}
\left( \delta \square \right) T^{\mu \nu \alpha \beta}=  \delta g^{\sigma \lambda} \nabla_\sigma \nabla_\lambda T^{\mu \nu \alpha \beta} + g^{\sigma \lambda} \bigg[ T^{\rho \nu \alpha \beta} \nabla_\sigma \delta \Gamma^\mu_{\lambda \rho} + \\
\\
+ T^{\mu \rho \alpha \beta} \nabla \delta \Gamma^\nu_{\lambda \rho} + T^{\mu \nu \rho \beta} \nabla_\sigma \delta \Gamma^\alpha_{\lambda \rho} + T^{\mu \nu \alpha \rho} \nabla_\sigma \delta \Gamma^\beta_{\lambda \rho} + \\
\\
+ 2 \delta \Gamma^\mu_{\lambda \rho} \nabla_\sigma T^{\rho \nu \alpha \beta} + 2 \delta \Gamma^\nu_{\lambda \rho} \nabla_\sigma T^{\mu \rho \alpha \beta} + 2 \delta \Gamma^\alpha_{\lambda \rho} \nabla_\sigma T^{\mu \nu \rho \beta}
+\\
\\
+ 2 \delta \Gamma^\beta_{\lambda \rho} \nabla_\sigma T^{\mu \nu \alpha \rho} - \delta \Gamma^\rho_{\sigma \lambda} \nabla_\rho T^{\mu \nu \alpha \beta}\biggl]\,.
\end{array}
\end{equation}

Thus, considering Eqs.(\ref{delta h3}) and (\ref{eq27}) after applying the symmetry properties of the Riemann tensor, we finally have

\begin{equation}\label{eq28}
\begin{array}{ll}
\int d^4x \sqrt{-g} R_{\mu \nu \alpha \beta} \delta h_3(-\square_\Lambda) R^{\mu \nu \alpha \beta}= \\
\\
-\frac{1}{\Lambda^2}\sum_{m=0}^{\infty} \sum_{r=0}^{m-1} h_3^{(m)} \int d^4x \sqrt{-g} \biggl\{  R^{[r]^\dagger}_{\pi \sigma \alpha \beta} \times\\
\\
\times \left( \nabla_\mu \nabla_\nu R^{[m-r-1] \pi \sigma \alpha \beta} \right) \delta g^{\mu \nu} + \biggl[  8 g^{\pi \mu} R^{[r]^\dagger}_{\alpha \omega \sigma \beta} \times \\
\\
\times \left( \nabla_\pi R^{[m-r-1] \nu \omega \sigma \beta}\right) - \frac{4}{\sqrt{-g}} \nabla_\pi \bigl( \sqrt{-g} g^{\pi \mu} R^{[r]^\dagger}_{\alpha \sigma \omega \beta} \times \\
\\
\times R^{[m-r-1] \nu \sigma \omega \beta} \bigr) - g^{\mu \nu} R^{[r]^\dagger}_{\pi \sigma \rho \beta} \left( \nabla_\alpha R^{[m-r-1] \pi \sigma \rho \beta} \right) \biggr]\delta \Gamma^\alpha_{\mu \nu}  \biggl\}\,,
\end{array}
\end{equation}
where, as usual, we have denoted

\begin{equation} \label{eq08}
R^{[k] \alpha \beta \sigma \omega} \equiv \left( -\frac{\square}{\Lambda^2}\right)^k R^{\alpha \beta \sigma \omega}\,,
\end{equation}

and

\begin{equation}\label{eq09}
R^{[k]^\dagger}_ {\alpha \beta \sigma \omega} \equiv \left( -\frac{\square^\dagger}{\Lambda^2}\right)^k R_{\alpha \beta \sigma \omega}\,.
\end{equation}

Finally, we have that the variation of the Riemann action leads to

\begin{equation}\label{functional derivative g riemann}
\begin{array}{ll}
\frac{\delta \mathcal{S}_{Riemann}}{\delta g^{\mu \nu}}= \sqrt{-g}
\biggl\{
-3 {{R^\pi}_{(\mu}}^{\alpha \beta} \, h_3 \left(-\frac{\square^\dagger}{\Lambda^2}\right)  R_{ \nu) \pi \alpha \beta} +\\
\\
- g_{\sigma (\mu}{R}_{\nu) \omega \alpha \beta} \,
h_3\left(-\frac{\square}{\Lambda^2} \right) R^{\sigma \omega
\alpha \beta}
- \frac{1}{2} g_{\mu \nu} R_{\pi \sigma \alpha \beta} \times \\
\\
\times  h_3\left(-\frac{\square}{\Lambda^2} \right)R^{\pi \sigma
\alpha \beta}
-\frac{1}{\Lambda^2} \sum_{m=0}^{\infty} \sum_{r=0}^{m-1} h_3^{(m)} R^{[r]^\dagger}_{\pi \sigma \alpha \beta} \times\\
\\
\times \nabla_{(\mu} \nabla_{\nu)} R^{[m-r-1] \pi \sigma \alpha
\beta}  \biggr\}\,,
\end{array}
\end{equation}

and

\begin{equation}\label{functional derivative Gamma riemann}
\begin{array}{ll}
\frac{\delta \mathcal{S}_{Riemann}}{\delta \Gamma^\alpha_{\mu
\nu}}= \nabla_\beta \biggl[ \sqrt{-g} g^{\rho (\nu} g^{\mu)
\sigma} g^{\beta \pi}
h_3\left(-\frac{\square^\dagger}{\Lambda^2} \right)  R_{\alpha \rho \sigma \pi} \biggr] +\\
\\
+  \nabla_\rho \biggr[\sqrt{-g} g_{\alpha \beta}
h_3\left(-\frac{\square}{\Lambda^2} \right)  R^{\beta (\nu \mu) \rho} \biggr] +\\
\\
-\frac{1}{\Lambda^2}\sum_{m=0}^{\infty} \sum_{r=0}^{m-1} h_3^{(m)}
\biggl\{ 8 g^{\pi (\mu} R^{[r]^\dagger}_{\alpha \omega \sigma
\beta}
\left[ \nabla_\pi R^{[m-r-1] \nu) \omega \sigma \beta}\right]+ \\
\\
- \frac{4}{\sqrt{-g}}\nabla_\pi \bigl[ \sqrt{-g} \, g^{\pi (\mu}
\, R^{[r]^\dagger}_{\alpha \sigma \omega \beta} R^{[m-r-1] \nu) \sigma \omega \beta} \bigr] +\\
\\
- g^{\mu \nu} R^{[r]^\dagger}_{\pi \sigma \rho \beta} \left[
\nabla_\alpha R^{[m-r-1] \pi \sigma \rho \beta} \right] \biggr\}\,.
\end{array}
\end{equation}

\section{Discussion}\label{section discussion}

In the last sections we have derived the functional derivatives of
the action of the non-local gravitational field with respect to
the metric tensor and the connection, see Eqs.(\ref{functional
derivative g R},\ref{functional derivative Gamma
R},\ref{functional derivative g ricci},\ref{functional derivative
Gamma ricci},\ref{functional derivative g riemann},\ref{functional
derivative Gamma riemann}). The dynamical equations for the metric
tensor and the connection read

\begin{equation}\label{Palatini eq 1}
\begin{array}{ll}
\frac{\delta \mathcal{S}_{grav}}{\delta g^{\mu \nu}}= \frac{\delta
\mathcal{S}_{EH}}{\delta g^{\mu \nu}} + \frac{\delta
\mathcal{S}_{Scalar}}{\delta g^{\mu \nu}} + \frac{\delta
\mathcal{S}_{Ricci}}{\delta g^{\mu \nu}} + \frac{\delta
\mathcal{S}_{Riemann}}{\delta g^{\mu \nu}} + \frac{\delta
\mathcal{S}_{m}}{\delta g^{\mu \nu}} = 0\,,
\end{array}
\end{equation}
and
\begin{equation}\label{Palatini eq 2}
\begin{array}{ll}
\frac{\delta \mathcal{S}_{grav}}{\delta \Gamma^\alpha_{\mu \nu}}=
\frac{\delta \mathcal{S}_{EH}}{\delta \Gamma^\alpha_{\mu \nu}} +
\frac{\delta \mathcal{S}_{Scalar}}{\delta \Gamma^\alpha_{\mu \nu}}
+ \frac{\delta \mathcal{S}_{Ricci}}{\delta \Gamma^\alpha_{\mu
\nu}} + \frac{\delta \mathcal{S}_{Riemann}}{\delta
\Gamma^\alpha_{\mu \nu}} + \frac{\delta \mathcal{S}_{m}}{\delta
\Gamma^\alpha_{\mu \nu}} = 0\,.
\end{array}
\end{equation}

At a first look Eqs.(\ref{Palatini eq 1},\ref{Palatini eq 2}) seem
extremely complicated. An initial  observation is that it is
impossible to recast (\ref{Palatini eq 2}) in the form (\ref{eq
einstein palatini 2b}), i.e. it is not immediate to find a metric
tensor $h_{\mu\nu}$ for which $\Gamma^\alpha_{\mu \nu}$ is the
metric connection. Therefore we lose one of the main
simplifications of the Palatini formalism in $f(R)$ theories.

However, with certain assumptions, one can show that some of the
well known solutions of the Einsteinian gravity verify
(\ref{Palatini eq 1},\ref{Palatini eq 2}). Let us set
$\mathcal{S}_m = 0$ and $h_3(\square_\Lambda)=0$, which guarantees
that  $ \delta \mathcal{S}_{Riemann}/\delta g_{\mu \nu} = 0$ and $
\delta \mathcal{S}_{Riemann}/\delta \Gamma^\alpha_{\mu \nu} = 0$.

As first example, let us assume that $\Lambda_c =0$  and let us
consider the vacuum solutions of the Einstein's equations, which
are such that $R_{\mu\nu}=0$, where the Ricci tensor is
constructed with the metric connection
$\Gamma^{(met)\alpha}_{\quad\quad \mu\nu}$ of $g_{\mu\nu}$, that
verifies $\nabla^{(met)}_\alpha g_{\mu\nu} =0$. It is easy to show
that the couple  ($\Gamma^{(met)\alpha}_{\quad\quad \mu\nu}$,
$g_{\mu\nu}$) is solution of the system (\ref{Palatini eq
1},\ref{Palatini eq 2}) in vacuum and with zero cosmological
constant. In fact, since $R=0$ and $R_{\mu \nu}=0$, one has that
$\delta \mathcal{S}_{EH}/\delta g^{\mu \nu} =0$, $\delta
\mathcal{S}_{Scalar}/\delta g^{\mu \nu}=0$, $ \delta
\mathcal{S}_{Ricci}/\delta g^{\mu \nu}=0$ independently. For the
same reason, using the fact that the connection is metric, one
also has   $\delta \mathcal{S}_{EH}/ \delta \Gamma^\alpha_{\mu
\nu} =0$, $\delta \mathcal{S}_{Scalar}/ \delta \Gamma^\alpha_{\mu
\nu}=0$, $ \delta \mathcal{S}_{Ricci}/\delta \Gamma^\alpha_{\mu
\nu}$, so that the system (\ref{Palatini eq 1},\ref{Palatini eq
2}) is fully verified.

A second class of exact solutions of (\ref{Palatini eq
1},\ref{Palatini eq 2}) is obtained considering a nonzero
cosmological constant $\Lambda_c$. One can thus show that the
couple ($\Gamma^{(met)\alpha}_{\quad\quad \mu\nu}$, $g_{\mu\nu}$)
of solutions of the Einstein's equations in vacuum with
cosmological constant, is a solution of (\ref{Palatini eq
1},\ref{Palatini eq 2}). In fact, in this case one has $R_{\mu\nu}
= - \Lambda_c \, g_{\mu\nu}/2$ and therefore
$\nabla^{(met)}_\alpha R_{\mu\nu} =0$.   One can easily verify
that  $\delta \mathcal{S}_{grav}/\delta g_{\mu \nu} =0$ and
$\delta \mathcal{S}_{grav}/\delta \Gamma^\alpha_{\mu \nu}=0$, so
that the equations of motion are satisfied.

These two cases exhaust the  known examples of exact solutions of
our dynamical equations. In more general cases one can include a
contribution (\ref{action3}) to the action of the gravitational
field, or extend the non-locality of the gravitational field to
the matter fields, as in the case of a non-local scalar field
briefly studied in appendix \ref{section nonlocal scalar}, which
implies $ \delta \mathcal{S}_{m}/ \delta \Gamma^\alpha_{\mu \nu}
\neq 0$.

These simple examples show that, at least in the case
$h_3(\square_\Lambda) =0$, the singularities of Einstein's gravity
are not removed, since for instance black hole solutions are still
there. However, one hopes that in the more general case
$h_3(\square_\Lambda) \neq 0$, models can be constructed which are
singularity free.

\section{Conclusions}\label{section conclusions}

In this paper we have found the dynamical equations of the
non-local gravity model (\ref{lagrangian density}) in the Palatini
formalism. We have discussed some of the properties of the model
and we have shown that, in the case $h_3=0$, the vacuum solutions
of general relativity are also vacuum solutions of the model
(\ref{lagrangian density}). We have concluded that, at least in
this case, the singularities of Einstein's gravity are not
removed.

\section*{Acknowledgements.}
MPL thanks CAPES for financial support.
\appendix

\section{Variation of the action (\ref{action4})}\label{section action 4}

 In this section, the goal will be to display the calculus of the variation of the action given by Eq.(\ref{action4}).
 Thus, after some straightforward manipulation considering Eq.(\ref{definition box dagger}) and expanding $h_2^*$ in power series similarly to Eq.(\ref{definition h2}),
 it is not difficult to show that the variation of the Ricci action can be written in the following manner

\begin{equation}\label{variation ricci*}
\begin{array}{ll}
\delta \mathcal{S}^*_{Ricci}= \int d^4x \sqrt{-g} \biggl \{
\biggl[-\frac{1}{2} g_{\mu \nu} R^{\alpha \beta}
h_2^*(-\frac{\square}{\Lambda^2}) R_{\alpha \beta} + \\
\\
+  R_{(\nu}^{\,\,\,\,\, \alpha}
h_2^*\left(-\frac{\square}{\Lambda^2} \right) R_{\mu) \alpha} +
R^\alpha_{\,\,\, (\nu} h_2^*\left(-\frac{\square}{\Lambda^2}
\right) R_{\alpha \mu)} \biggr]\delta g^{\mu \nu}
+ \\
\\
+ \biggl[ g^{\alpha \mu} g^{\beta \nu} \biggl( h_2^*\left(-\frac{\square}{\Lambda^2}\right) R_{\alpha \beta} \biggr) + \biggl( h_2^*\bigl( -\frac{\square^\dagger}{\Lambda^2}\bigr) R^{\mu \nu}\biggr)\biggr] \delta R_{\mu \nu} + \\
\\
+R^{\mu \nu} \delta
h_2^*\left(-\frac{\square}{\Lambda^2}\right)  R_{\mu \nu}
\biggr\}\,.
\end{array}
\end{equation}

By considering (\ref{delta ricci tensor}), the terms in the second bracket on the r.h.s. of (\ref{variation ricci*}) can be handled to give

\begin{equation}
\begin{array}{ll}
\int d^4x \sqrt{-g} \delta R_{\mu \nu} \biggl[ g^{\alpha \mu} g^{\beta \nu} h_2^*\left(-\frac{\square}{\Lambda^2}\right) R_{\alpha \beta} +
h_2^*\bigl( -\frac{\square^\dagger}{\Lambda^2}\bigr) R^{\mu \nu} \biggr]=\\
\\
\int d^4x \biggl[ \nabla_\rho \left( \sqrt{-g}
\delta^{(\nu}_\alpha F^{*\mu) \rho} \right) - \nabla_\alpha \left(
\sqrt{-g} F^{*(\nu \mu)} \right) \biggr] \delta \Gamma^\alpha_{\mu
\nu}\,,
\end{array}
\end{equation}
where $F^{* \alpha \beta} \equiv g^{\omega \alpha} g^{\pi \beta}
h_2^*\left( -\frac{\square}{\Lambda^2}\right) R_{\omega \pi} +
h_2^*\left( -\frac{\square^\dagger}{\Lambda^2}\right) R^{\alpha
\beta}$.

With the aim of treating the last term on the r.h.s. of (\ref{variation ricci*}),
by using (\ref{delta box generic}) an expression analogous to (\ref{eq12}) can be found. In this case, it is given by

\begin{equation}\label{eq16}
\begin{array}{ll}
 \left( \delta \square\right) T_{\alpha \beta}=
\delta g^{\mu \nu}\nabla_\mu
\nabla_\nu T_{\alpha \beta}- T_{\sigma \beta}  \nabla^\mu \delta \Gamma^\sigma_{\mu
\alpha} -  T_{\alpha \sigma}  \nabla^\mu \delta \Gamma^\sigma_{\mu \beta} +\\
\\
- g^{\mu \nu} \delta \Gamma^\lambda_{\mu \nu} \nabla_\lambda
T_{\alpha \beta} - 2 \delta\Gamma^\sigma_{\mu \alpha} \nabla^\mu T_{\sigma \beta} -
2  \delta \Gamma^{\sigma}_{\mu \beta} \nabla^\mu T_{\alpha \sigma}\,,
\end{array}
\end{equation}
with $T_{\alpha \beta}$ denoting an arbitrary covariant tensor of rank $2$.

Thereby, taking into account Eq.(\ref{eq16}) and that here we have
\begin{equation}
\delta h_2^* \left(-\square_\Lambda\right)= \sum^\infty_{m=0} \sum^{m-1}_{r=0} \frac{h^{*(m)}_2}{\left(-\Lambda^2\right)^m}
\square^r \delta \square \, \, \square^{m-r-1}\,,
\end{equation}
we write

\begin{equation}\label{eq17}
\begin{array}{ll}
\int d^4x \sqrt{-g}  R^{\mu \nu} \delta h_2^*
\left(-\square_\Lambda \right) R_{\mu \nu}=
-\frac{1}{\Lambda^2} \sum_{m=0}^{\infty} \sum_{r=0}^{m-1} h_2^{*(m)} \times\\
\\
\int d^4x \biggl\{ \biggl(  \sqrt{-g}  R^{[r]^\dagger \alpha \beta} \nabla_{(\mu} \nabla_{\nu)} R^{[m-r-1]}_{\alpha \beta} \biggr) \delta g^{\mu \nu} +\\
\\
+ \biggl[ \nabla_\rho \biggl( \sqrt{-g} g^{\mu \rho} \left(
R^{[r]^\dagger \nu \beta} R^{[m-r-1]}_{\alpha \beta} +
R^{[r]^\dagger \beta \nu} R^{[m-r-1]}_{\beta \alpha}\biggr)\right)+\\
\\
- \sqrt{-g}\left( g^{\mu \nu}  R^{[r]^\dagger \rho \beta}
\nabla_\alpha R^{[m-r-1]}_{\rho \beta} +  2 R^{[r]^\dagger \nu
\beta} \nabla^\mu R^{[m-r-1]}_{\alpha \beta}+\right.\\
\\
\left.+2 R^{[r]^\dagger \beta \nu} \nabla^\mu R^{[m-r-1]}_{ \beta
\alpha} \right) \biggr] \delta \Gamma^\alpha_{\mu \nu} \biggr\}\,,
\end{array}
\end{equation}

where we have defined

\begin{equation}
R^{[k]}_{\alpha \beta} \equiv \left( -\frac{\square}{\Lambda^2}\right)^k R_{\alpha \beta}\,,
\end{equation}

and

\begin{equation}
R^{[k]^\dagger \alpha \beta} \equiv \left( -\frac{\square^\dagger}{\Lambda^2}\right)^k R^{\alpha \beta}\,.
\end{equation}

Therefore, the complete variation of the Ricci action (\ref{action4}) reads
\begin{equation}
\begin{array}{ll}
\frac{\delta \mathcal{S}^*_{Ricci}}{\delta g^{\mu \nu}}= \sqrt{-g}
\biggl[-\frac{1}{2} g_{\mu \nu} R^{\alpha \beta}
h_2^*(-\frac{\square}{\Lambda^2}) R_{\alpha \beta} + \\
\\
+  R_{(\nu}^{\,\,\,\,\, \alpha}
h_2^*\left(-\frac{\square}{\Lambda^2} \right) R_{\mu) \alpha} +
R^\alpha_{\,\,\, (\nu} h_2^*\left(-\frac{\square}{\Lambda^2}
\right) R_{\alpha \mu)} +\\
\\
-\frac{1}{\Lambda^2} \sum_{m=0}^{\infty} \sum_{r=0}^{m-1}
h_2^{*(m)} R^{[r]^\dagger \alpha \beta}  \nabla_{(\mu}
\nabla_{\nu)} R^{[m-r-1]}_{\alpha \beta} \biggr]\,,
\end{array}
\end{equation}

and
\begin{equation}
\begin{array}{ll}
\frac{\delta \mathcal{S}^*_{Ricci}}{\delta \Gamma^\alpha_{\mu
\nu}}= \nabla_\rho \left( \sqrt{-g} \delta^{(\nu}_\alpha F^{*\mu)
\rho} \right) - \nabla_\alpha \left(
\sqrt{-g} F^{*(\nu \mu)} \right) +\\
\\
-\frac{1}{\Lambda^2} \sum_{m=0}^{\infty} \sum_{r=0}^{m-1}
h_2^{*(m)} \times\\
\\
\biggl\{ \nabla_\rho \biggl[ \sqrt{-g} g^{\rho(\mu } \left(
R^{[r]^\dagger \nu) \beta} R^{[m-r-1]}_{\alpha \beta} +
R^{[r]^\dagger \beta \nu)} R^{[m-r-1]}_{\beta \alpha}\biggr)\right]+\\
\\
- \sqrt{-g}\left[ g^{\mu \nu}  R^{[r]^\dagger \rho \beta}
\nabla_\alpha R^{[m-r-1]}_{\rho \beta} +  2 R^{[r]^\dagger (\nu
\beta} \nabla^{\mu)} R^{[m-r-1]}_{\alpha \beta}+\right.\\
\\
\left.+2 R^{[r]^\dagger \beta (\nu} \nabla^{\mu)} R^{[m-r-1]}_{
\beta \alpha} \right] \biggr\}
\end{array}
\end{equation}
where the indices $\mu$ and $\nu$ have been symmetrized.

\section{Non-local scalar field}\label{section nonlocal scalar}

Let us consider a non-local scalar field whose action is

\begin{equation}\label{phi action}
\begin{array}{ll}
\delta \mathcal{S}_{\phi}= \int d^4x \sqrt{-g} \biggl\{ \phi \,
Q\left(-\square_\Lambda\right)  \phi - V(\phi) \biggr\}
 \, ,
\end{array}
\end{equation}
where $Q(z)$ is analytic, and let us calculate the functional
derivatives of this action with respect to the metric and the
connection. The variation of (\ref{phi action}) gives

\begin{equation}\label{delta phi 1}
\begin{array}{ll}
\delta \mathcal{S}_{\phi}= \int d^4x \sqrt{-g} \biggl\{ \phi \,
\delta Q\left(-\square_\Lambda\right)  \phi + \\
\\
+ \frac{V(\phi)- \phi \, Q\left(-\square_\Lambda\right)  \phi}{2}
g_{\mu\nu} \delta g^{\mu\nu}\biggr\} \, ,
\end{array}
\end{equation}
therefore all wee need to calculate is the first term in the
r.h.s. of Eq.(\ref{delta phi 1}). To do that, we follow the same
procedure of section \ref{section action 1}, and we expand
$Q\left(-\square_\Lambda\right)$ in power series as

\begin{equation} \label{definition h1 Q}
Q(-\square_\Lambda) = \sum^\infty_{m=0} Q^{(m)}
\left(-\frac{\square}{\Lambda^2}\right)^m \, ,
\end{equation}
which allows to express $\delta Q\left(-\square_\Lambda\right)$ as

\begin{equation}\label{delta h1 Q}
\delta Q\left(-\square_\Lambda\right)= \sum^\infty_{m=0}
\sum^{m-1}_{r=0} \frac{Q^{(m)}}{\left(-\Lambda^2\right)^m}
\square^r \delta \square \, \, \square^{m-r-1} \, ,
\end{equation}
so that by use of (\ref{delta box}) one has

\begin{equation}\label{delta phi b}
\begin{array}{ll}
\int d^4 x \sqrt{-g} \phi \delta Q\left(-\square_\Lambda\right) \phi= - \frac{1}{\Lambda^2} \sum^\infty_{m=0} \sum^{m-1}_{r=0} Q^{(m)} \times \\
\\
\times \int d^4 x \sqrt{-g}
\biggl[ \left( \phi^{[r]^\dagger} \nabla_\mu \nabla_\nu \phi^{[m-r-1]} \right) \delta g^{\mu \nu} + \\
\\
- \left(  \phi^{[r]^\dagger} g^{\mu \nu} \nabla_\alpha
\phi^{[m-r-1]} \right) \delta \Gamma^\alpha_{\mu \nu} \biggr]\, ,
\end{array}
\end{equation}
where we have defined
\begin{equation} \label{eq21-b}
\phi^{[k]} \equiv \left( -\frac{\square}{\Lambda^2}\right)^k \phi\,,
\end{equation}
and
\begin{equation}\label{eq22-b}
\phi^{[k]^\dagger} \equiv \left(
-\frac{\square^\dagger}{\Lambda^2}\right)^k \phi \,.
\end{equation}

Therefore, from Eqs.(\ref{delta phi 1},\ref{delta phi b}) one has

\begin{equation}\label{EoM phi 1}
\begin{array}{ll}
\frac{\delta \mathcal{S}_{\phi}}{\delta g^{\mu \nu}} =\sqrt{-g}
\biggl\{  \frac{V(\phi)- \phi \, Q\left(-\square_\Lambda\right)
\phi}{2} g_{\mu\nu} +\\
\\
- \frac{1}{\Lambda^2} \sum^\infty_{m=0} \sum^{m-1}_{r=0} Q^{(m)}
\, \phi^{[r]^\dagger} \nabla_{(\mu} \nabla_{\nu)} \phi^{[m-r-1]}
\biggr\}\, ,
\end{array}
\end{equation}
and

\begin{equation}\label{EoM phi 2}
\begin{array}{ll}
\frac{\delta \mathcal{S}_{\phi}}{\delta \Gamma^\alpha_{\mu \nu}} =
\frac{\sqrt{-g}}{\Lambda^2} \sum^\infty_{m=0} \sum^{m-1}_{r=0}
Q^{(m)} \,  \phi^{[r]^\dagger} g^{\mu \nu} \nabla_\alpha
\phi^{[m-r-1]} \, .
\end{array}
\end{equation}

\end{document}